\documentclass{elsart}

\usepackage{epsfig}

\usepackage{amssymb}

\begin{document}

\begin{frontmatter}

\title{Effect of pressure on the superconducting transition temperature of doped and neutron-damaged MgB$_2$}

\author{S. L. Bud'ko, R. H. T. Wilke, M. Angst, and P. C. Canfield}

\address{Ames Laboratory US DOE and Department of Physics and Astronomy, Iowa State University, Ames, IA 50011}

\begin{abstract}
Measurements of the superconducting transition temperatures for Al-doped, C-doped and neutron-damaged-annealed
MgB$_2$ samples under pressure up to $\sim 8$ kbar are presented. The $dT_c/dP$ values change systematically with
the decrease of the ambient pressure $T_c$ in a regular fashion. The evolution of the pressure derivatives can be
understood assuming that the change in phonon spectrum is a dominant contribution to $dT_c/dP$.
\end{abstract}

\end{frontmatter}

The recently discovered \cite{1} superconductivity, with a superconducting transition temperature, $T_c \sim 40$
K, in the simple binary intermetallic compound, MgB$_2$, attracted a lot of attention to this material. Over the
past few years a great deal of knowledge (phonon-mediated mechanism of superconductivity, high,
temperature-dependent anisotropy of the upper critical field, two-gap nature of the superconducting state, {\it
etc.}) was obtained about the pure MgB$_2$ samples (see {\it e.g.} \cite{2,3,4,5} for reviews). One of the
approaches frequently used in attempts to understand the mechanism of superconductivity in a particular material
and, when possible, to improve its superconducting properties, is to study material's response to a small
perturbation, like chemical substitution and/or pressure/strain. Both substitution and pressure effects on $T_c$
of magnesium diboride were studied for a few years by a number of groups. Apparently the substitutional chemistry
of MgB$_2$ is very limited, with only carbon (replacing boron) and aluminum (replacing magnesium) being
unambiguous dopants \cite{6}. Pressure studies were performed so far only on nominally pure MgB$_2$ (including
samples with isotopically enriched $^{10}$B and $^{11}$B), but even for this case a substantial spread of the
T$_c$ values at ambient pressure ($\sim 35$ K to 40.5 K) (a spread which reveals that some of the samples were
heavily defected and/or unintentionally doped, during the synthesis), as well as different pressure media used,
still caused some discussion as to the exact, intrinsic, pressure derivative ($dT_c/dP$) value and the origin of
the observed differences (see {\it e.g.} \cite{7,8}).

In this work we extend the available pressure data to samples for which $T_c$ is intentionally and significantly
suppressed by different means in an attempt to have a broader picture of the physical properties of magnesium
diboride.

Several types of samples were used for this study. The pure MgB$_2$ sample was in the form of wire filaments,
similar to those described in \cite{9}. Carbon-doped samples, Mg(B$_{1-x}$C$_x$)$_2$ were: wire filaments
synthesized in the manner described in \cite{10} from the filaments on which boron and carbon were co-deposited by
a chemical vapor deposition process \cite{11} (for carbon levels up to $\sim 3.8$\% C); bulk
Mg(B$_{0.9}$C$_{0.1}$)$_2$ synthesized from B$_4$C and elemental magnesium as described in \cite{12}, and bulk
Mg(B$_{0.925}$C$_{0.075}$)$_2$ synthesized from B$_4$C, boron and magnesium by two-step vapor diffusion and
solid-state reaction \cite{13}. The carbon content in the filaments as well as in the latter bulk sample was
estimated based on the shift of the {\it a} lattice parameter following the procedure described in \cite{10} based
on the results of \cite{14}. The bulk aluminum-doped sample, with a nominal chemical formula
Mg$_{0.8}$Al$_{0.2}$B$_2$, was synthesized from elemental magnesium, aluminum and boron by two-step vapor
diffusion and solid-state reaction \cite{15}. Ambient pressure superconducting transition temperatures for all
samples were consistent with the published data for the respective materials. Another group of the samples used
for this study consisted of neutron-irradiated (fluence of $\sim 10^{19}$ n/cm$^2$) and annealed MgB$_2$
filaments. The damaged filaments were annealed for 24 hours at different temperatures and had superconducting
transitions between $\sim 8$ K and $\sim 30$ K \cite{16}. The choice of the samples for this work offers (i) very
broad range of the ambient pressure $T_c$ values, from $\sim 8$ K to $\sim 40$ K; (ii) possibility to reach
similar transition temperatures by different means ({\it e.g.} $T_c \approx 22 - 23$ K at ambient pressure was
obtained by carbon doping in Mg(B$_{0.9}$C$_{0.1}$)$_2$, by aluminum doping in Mg$_{0.8}$Al$_{0.2}$B$_2$, and by
annealing the neutron-damaged filaments at 300$^\circ$C for 24 hours).

\begin{figure}
\begin{center}
\includegraphics[angle=0,width=120mm]{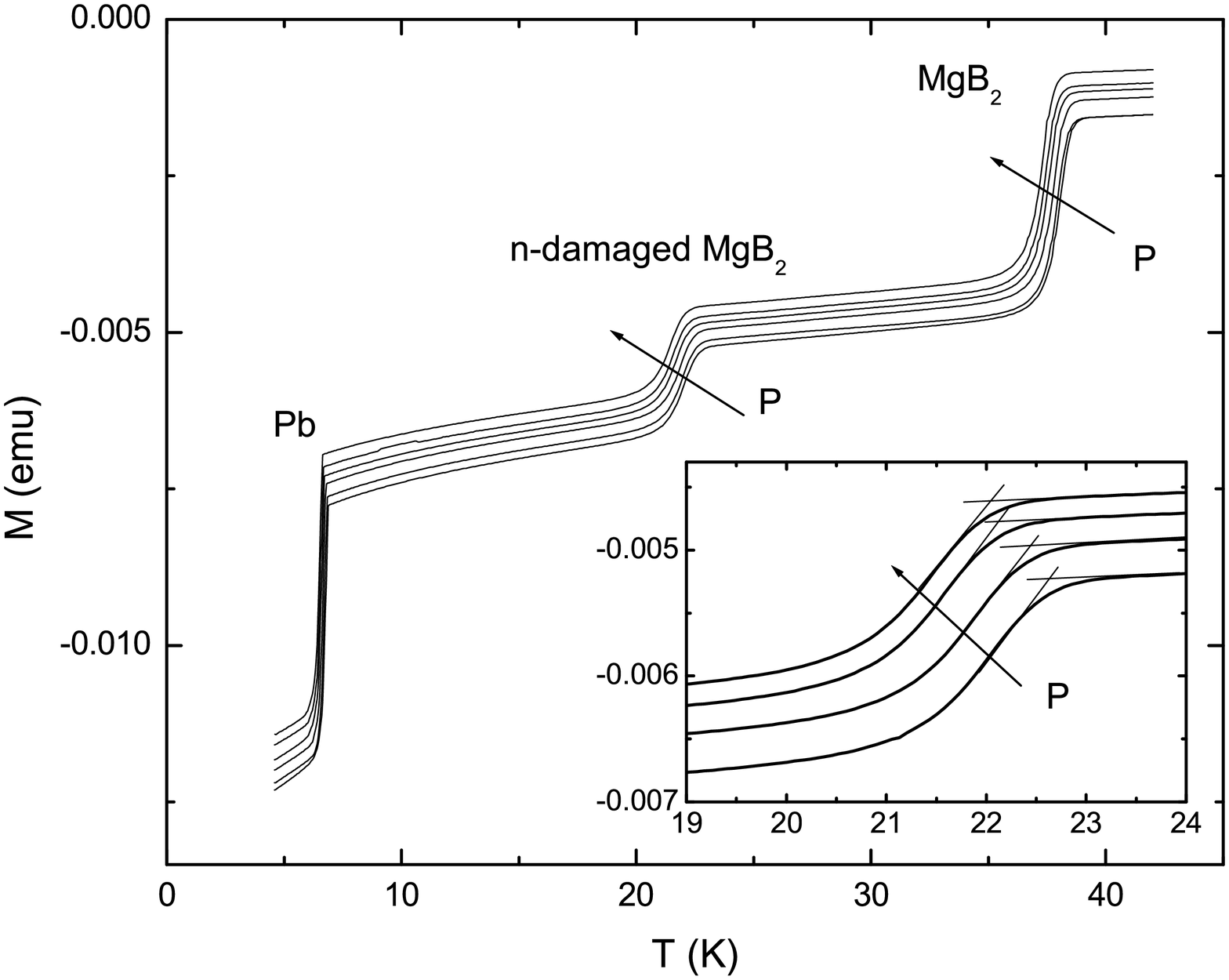}
\end{center}
\caption{Representative magnetization curves for one of the experiments. The pressure cell load for this
experiment was: Pb (pressure gauge), neutron-damaged MgB$_2$ filaments annealed at $300^\circ$C for 24 hours and
pure, undamaged MgB$_2$ filaments.  Pressure values for the different curves are: 1.2, 2.3, 4.0, 5.6, 6.7, 8.1
kbar (arrows indicate the direction of pressure increase). Inset: enlarged part of a subset of the curves
corresponding to superconducting transition in the neutron damaged and annealed sample. Pressure values are 1.2,
4.0, 6.7 and 8.1 kbar ($P$ increases in the direction of the arrow). The $T_c$ criterion (onset) is shown by the
intercept of the straight lines for each curve.}\label{f1}
\end{figure}

The piston-cylinder clamp-type pressure cell used in this work was designed to fit a commercial Quantum Design
MPMS-5 SQUID magnetometer. It was made out of non-magnetic Ni-Co alloy {\it MP35N} (see \cite{17} for a discussion
of the use of the {\it MP35N} alloy for pressure cells). The dimensions of the body of the cell were 8.3 mm O.D.,
4 mm I.D., 150 mm length. Pressure was generated in a teflon capsule filled with approximately 50:50 mixture of
n-pentane and mineral oil \cite{18}. The shift in the superconducting transition temperature of 6N purity Pb was
used to determine pressure at low temperatures \cite{19} (below $\sim 75$ K change of pressure caused by
differences in the thermal expansion of the pressure cell and the pressure media is insignificant \cite{19a}). The
design of the cell allows for the routine establishment of pressures in excess of 8 kbar at low temperatures. DC
magnetization measurements were performed in an applied field of 50 Oe in a zero-field-cooled warming protocol.
The measurements were performed in such a mode that temperature was nominally stable at each data point
measurement; to reduce possible gradients the average warming rate did not exceed 0.05 K/min. To save on the total
measurements time as well as to have an additional caliper for the differences in the pressure responses for the
different samples, the lead pressure gauge and two samples (chosen in such a way that their $T_c$ values were well
separated from each other) were usually measured simultaneously (Fig. \ref{f1}). An onset criterion was used to
determine the superconducting transition temperatures (Fig. \ref{f1}, inset). There was no appreciable broadening
of the transitions under pressure.

\begin{figure}
\begin{center}
\includegraphics[angle=0,width=80mm]{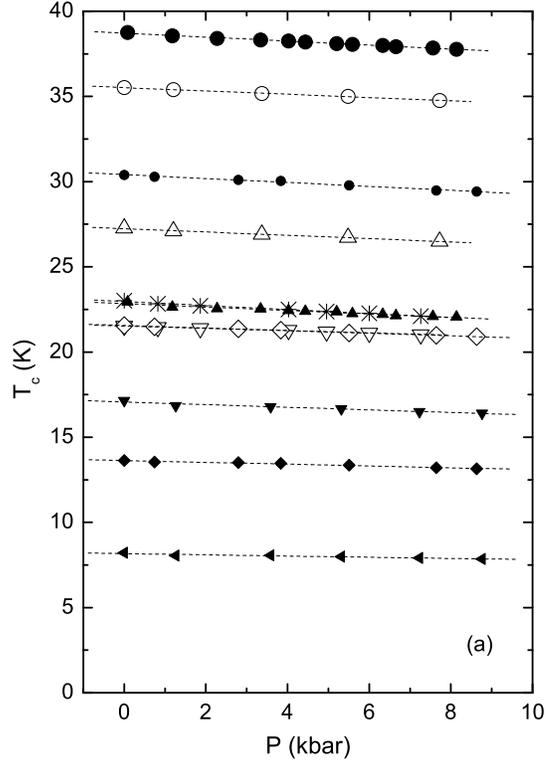}
\includegraphics[angle=0,width=80mm]{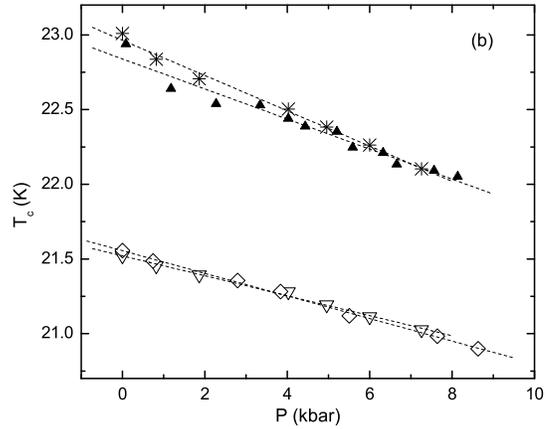}
\end{center}
\caption{(a) Pressure-dependent superconducting transition temperatures for the samples used in this work: pure
MgB$_2$ (large filled circles);  Mg$_{0.8}$Al$_{0.2}$B$_2$ (asterisks); carbon-doped MgB$_2$ (open symbols:
circles - 3.8\%C, up-triangles - 7.5\%C, down-triangles and diamonds - two different batches of 10\%C);
neutron-damaged and annealed for 24 hours MgB$_2$ (small filled symbols: circles - $400{^\circ}$C, up-triangles -
$300{^\circ}$C, down-triangles - $250{^\circ}$C, diamonds - $200{^\circ}$C, left-triangles - $150{^\circ}$C).
Dashed lines - linear fits. (b) Enlarged part of the graph with the data for Mg$_{0.8}$Al$_{0.2}$B$_2$,
neutron-damaged and annealed at $300{^\circ}$C, and Mg(B$_{0.9}$C$_{0.1}$)$_2$ samples.}\label{f2}
\end{figure}

\begin{figure}
\begin{center}
\includegraphics[angle=0,width=120mm]{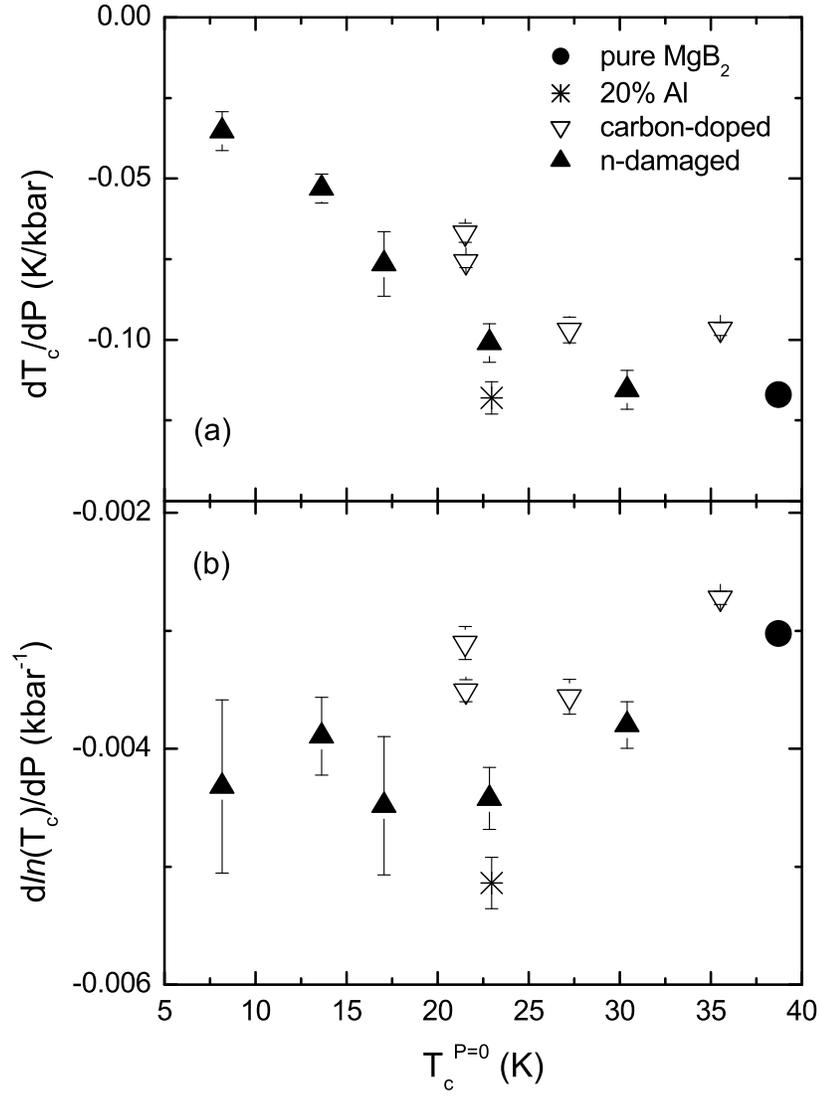}
\end{center}
\caption{Pressure derivatives, $dT_c/dP$ (panel (a)) and $d\ln T_c/dP$ (panel (b)), as a function of the
ambient-pressure $T_c$. Symbols: filled circle - pure MgB$_2$, asterisk - Mg$_{0.8}$Al$_{0.2}$B$_2$, open
down-triangles - carbon doped samples, filled up-triangles - neutron-damaged and subsequently annealed
samples.}\label{f3}
\end{figure}

Changes in the superconducting transition temperatures under pressure for all of the samples used in this study
are shown in Fig. \ref{f2}(a). In the pressure range ($P < 9$ kbar) discussed here $T_c$ values for all samples
decrease with pressure in a linear fashion without any apparent anomalies. The pressure derivatives, $dT_c/dP$ as
well as $d \ln T_c/dP$ are shown in Fig. \ref{f3}. For pure MgB$_2$ our value of $dT_c/dP = -0.117 \pm 0.003$
K/kbar is very close to the accepted  values for pure material under hydrostatic pressure \cite{8}. Both, for
carbon-doped and for neutron-damaged-annealed samples the absolute pressure derivative, $dT_c/dP$, is lower (to a
different extent for two aforementioned groups of samples) in the absolute value for the samples that have lower
ambient pressure transition temperature, $T_{c0}$. This trend is opposite to the one reported in \cite{7}, where
$\sim 37$ K sample had 2 times {\it higher} $|dT_c/dP|$ than the ''standard'' $\sim 39$ K sample. The relative
pressure derivative, $d\ln T_c/dP = 1/T_{c0} \times dT_c/dP|_{P=0}$ increases in the absolute value for samples
with lower $T_{c0}$ and, for neutron-damaged-annealed samples with $T_c \leq 23$ K levels to $\approx -0.0043$
kbar$^{-1}$. There seems to be a resolvable quantitative difference in pressure response for the MgB$_2$ samples
modified by carbon-doping, neutron damage and aluminum-doping (Fig. \ref{f3}). The simplest way to see this
difference is to compare the samples with similar values of $T_{c0}$. Within the set of materials studied, three
samples, Mg(B$_{0.9}$C$_{0.1}$)$_2$ ($T_{c0} = 21.6$ K), Mg$_{0.8}$Al$_{0.2}$B$_2$ ($T_{c0} = 23.0$ K) and
neutron-damaged filaments annealed at 300$^\circ$C for 24 hours ($T_{c0} = 22.8$ K), have significantly depressed,
yet similar, superconducting transition temperature (to $\sim 1/2 T_{c0}$ of pure MgB$_2$). The pressure
derivatives $dT_c/dP$ are $-0.071 \pm 0.005$, $-0.118 \pm 0.005$, and $-0.101 \pm 0.006$ K/kbar respectively (Fig.
\ref{f2}(b)), up to factor of 1.5 different. This simple comparison shows that $T_{c0}$ is not the sole factor
that determines $dT_c/dP$ of modified MgB$_2$ superconductors.

Band structure calculations for pure and doped MgB$_2$ (see \cite{20,21,22} and references therein for a review)
have proposed several parameters affecting the value of $T_c$ in these materials: density of states at the Fermi
level (DOS) ; frequency of the salient phonon mode ($E_{2g}$) and the anharmonicity of the phonon spectra.
Initially, let us assume that the (anisotropic) compressibilities of the samples used in this study are very close
to that for pure MgB$_2$. Within the rigid band approximation, the Mg(B$_{0.9}$C$_{0.1}$)$_2$ and
Mg$_{0.8}$Al$_{0.2}$B$_2$ samples will have the same shift of the Fermi level caused by doping, so the
contribution from the change in the DOS to the $dT_c/dP$ is expected to be the same for both compounds. The
position and width of the $E_{2g}$ mode as seen in Raman spectra (see {\it e.g.} \cite{23,24}) change somewhat
differently with carbon and aluminum doping, that allows us to infer that the difference in phonon spectra and in
anharmonicity of the relevant mode is a likely source of the different $dT_c/dP$ values. This conclusion fits well
within the general view \cite{8,25,26} that the electron-phonon interaction rather than the change in the density
of states determines pressure derivative of $T_c$ in MgB$_2$. Furthermore, one can also make a very loose,
qualitative, argument that since the $E_{2g}$ mode is associated with boron vibrations and Al-doping (affecting
primarily the Mg-sheet), neutron damage (affecting supposedly both sheets) and C-doping (mainly affecting B-sheet)
have an increasing effect on boron-layer, the pressure derivatives $dT_c/dP$ should be closer to that for pure
MgB$_2$ in the case of Al-doping and further from the MgB$_2$ value in the case of C-doping, the progression
observed in the experiment. To address the validity of these simple arguments, Raman measurements under pressure
for the three different 22-23 K samples as well as detailed band structure calculations of the pressure response
of $T_c$ to doping are desirable.

In summary, the pressure derivatives of the superconducting transition temperatures of Al-doped, C-doped and
neutron-damaged-annealed MgB$_2$ samples were measured up to $\sim 8$ kbar. For all samples the pressure
derivative remains negative. For C-doped and neutron-damaged-annealed samples $dT_c/dP$ decreases in absolute
value for samples with  smaller ambient pressure $T_{c0}$ values. Differences in $dT_c/dP$ for differently
modified samples with similar $T_{c0}$ values can be qualitatively understood by different levels of effect of
impurities/damage on boron layers and therefore on details of the phonon spectra.

{\it Acknowledgments:} SLB and PCC thank Scott Hannahs for suggesting {\it MP35N} as a possible cell material. We
appreciate help of John Farmer in neutron-irradiation of the samples. Ames Laboratory is operated for the U.S.
Department of Energy by Iowa State University under Contract No. W-7405-Eng-82. This work was supported by the
Director for Energy Research, Office of Basic Energy Sciences.

\clearpage

\clearpage

\clearpage

\end{document}